# High-resolution imaging of the gravitational lens candidate 1208+1011 with the Nordic Optical Telescope*

**Jens Hjorth**[1,**], **Frank Grundahl**[1], **Kari Nilsson**[2,3], **and Leif Festin**[2,4]

[1] Institute of Physics and Astronomy, University of Aarhus, DK–8000 Århus C, Denmark
[2] Nordic Optical Telescope, Observatorio Roque de los Muchachos, Apartado 474, E–38700 Santa Cruz de La Palma, Spain
[3] Tuorla Observatory, SF–21500 Piikkiö, Finland
[4] Astronomiska Observatoriet i Uppsala, S–75120 Uppsala, Sweden

**Abstract.** The large-redshift ($z = 3.8$) quasar 1208+1011 has recently been discovered to be a gravitational lens candidate with a separation of 0.″47 between the two imaged components (A and B). *NOT* (*Nordic Optical Telescope*) and *HST* (*Hubble Space Telescope*) studies from 1992 probing primarily the continuum light show that the amplification of A relative to B is almost achromatic with the canonical value A:B≈4±0.1. In this paper we present high-resolution optical images (FWHM ≈ 0.″4–0.″5) from 1993 of the quasar. From our narrow-band CCD frames centred on the redshifted Ly$\alpha$/N V line at ∼5900 Å we find that A:B=3.85±0.2. Our broad-band $I$ images uncontaminated by line-emission are found to be affected by PSF variation over a range of 20″. However, by iteratively determining the PSF from the images themselves we show that the data can be accounted for by two point sources with A:B=3.35±0.2. These results imply that the continuum intensity ratio has decreased during 1992–1993 and that the amplification of the emission-line regions relative to the continuum emitting regions is different in the two components (under the gravitational lens hypothesis). The most conservative interpretation of these results is that 1208+1011 is a gravitationally lensed quasar in which component B is being microlensed, but the possibility that 1208+1011 is a binary quasar is not excluded by the present data.

## 1. Introduction

With a redshift of 3.8 the quasar 1208+1011 in 1986 represented the most distant known object in the universe (Hazard, McMahon & Sargent 1986; Sargent et al. 1986). Recently, it was found independently in the ESO Gravitational Lens Key Project (Magain et al. 1992a) and in the *Hubble Space Telescope (HST)* Snapshot Survey (Maoz et al. 1992) that this quasar actually consists of two images with a brightness ratio of A:B≈4 separated by only 0.″47. The existence of this close quasar pair can be attributed to gravitational lensing of a single quasar by a foreground galaxy (Schneider, Ehlers & Falco 1992; Refsdal & Surdej 1994). This hypothesis was reinforced by subsequent *HST* imagery (Bahcall et al. 1992a) and spectroscopy (Bahcall et al. 1992b) which showed that the spectra of the two components are very similar. Additional photometry is reported in Hjorth & Jensen (1993). Thus 1208+1011 is one of the most distant gravitational lens candidates ($z = 3.8$) with one of the smallest separations between its two components (0.″47). Alternatively, it may be a binary quasar (Magain et al. 1992a). The quasar is undetected in the radio at a level of 0.6 mJy (Maoz et al. 1992).

Despite the striking similarity of the *HST* spectra of the two components a few loose ends remain: (i) There are indications in the *HST* spectra that the blue parts of the redshifted Ly$\alpha$/N V line near 5900 Å and the strong O VI/Ly$\beta$ emission line near 5000 Å are slightly more strongly amplified in the A image than in the B image compared to the continuum value of 4. (ii) No lensing galaxy has been detected.

To address these issues we undertook a study with the 2.56 m *Nordic Optical Telescope* (*NOT*) on La Palma by obtaining high-resolution direct images in a narrow band centred on the redshifted Ly$\alpha$ line and in broad-band $I$. The results of these observations are reported in this paper. We find no convincing evidence for the lensing galaxy, but confirm that the Ly$\alpha$/N V emission-line

*Send offprint requests to*: J. Hjorth (Cambridge)
* Based on observations obtained with the *Nordic Optical Telescope* (*NOT*), La Palma
** *Present address*: Institute of Astronomy, Madingley Road, Cambridge CB3 0HA, UK; jens@mail.ast.cam.ac.uk

**Table 1.** Observing log, April 11 1993

| Image # | Filter | Exposure time (sec) | Airmass | UTC | FWHM[a] |
|---|---|---|---|---|---|
| 1 | 5903/88 | 600 | 1.07 | 00:48 | 0.″42 |
| 2 | 5903/88 | 600 | 1.09 | 01:02 | 0.″43 |
| 3 | 5903/88 | 600 | 1.10 | 01:16 | 0.″44 |
| 4 | 5903/88 | 600 | 1.12 | 01:25 | 0.″46 |
| 5 | 5903/88 | 600 | 1.15 | 01:42 | 0.″47 |
| 6 | 5903/88 | 600 | 1.17 | 01:54 | 0.″54 |
| 7 | 5903/88 | 600 | 1.21 | 02:06 | 0.″62 |
| 8 | 5903/88 | 600 | 1.24 | 02:19 | 0.″57 |
| 9 | *I* | 300 | 1.38 | 02:54 | 0.″47 |
| 10 | *I* | 300 | 1.42 | 03:01 | 0.″50 |
| 11 | *I* | 300 | 1.46 | 03:08 | 0.″48 |
| 12 | *I* | 300 | 1.50 | 03:15 | 0.″55 |
| 13 | *I* | 300 | 1.56 | 03:23 | 0.″60 |
| 14 | *I* | 300 | 1.61 | 03:30 | 0.″48 |
| 15 | *I* | 300 | 1.66 | 03:37 | 0.″53 |
| 16 | *I* | 300 | 1.73 | 03:45 | 0.″58 |

[a] Approximate values obtained by fitting a sum of two two-dimensional Gaussians to the PSF star

is more strongly amplified in A than in B relative to the continuum.

## 2. Observations and image analysis

The quasar has redshifted Ly$\alpha$/N V emission at ∼5900 Å (see the medium-resolution spectrum of Steidel (1990)). By observing in a narrow band centred on this line we aimed at examining whether the intensity ratio between the two components is the same as in the continuum. For these observations we used a narrow-band filter with a central wavelength of 5903 Å and a width of 88 Å FWHM. With the aim of looking for the lensing galaxy we used a broad-band *I* filter which had more than 50 % of the peak transmission between 7630 Å and 8900 Å, i.e., centred on the continuum part of the spectrum between the C IV and the Al III/C III] emission lines (Sargent et al. 1986). Thus, these images also serve as continuum comparison images.

We used the Stockholm CCD camera with a TK 512×512 chip at the Cassegrain focus of the *NOT* on April 11 1993 with a focal expander consisting of a single negative lens yielding a pixel scale of around 0.″1 pixel$^{-1}$, slightly dependent on wavelength, and a field of view of 50″×50″. From comparison with *HST* images we determined the exact pixel scale of the CCD to be 0.″1012 ± 0.″0003 pixel$^{-1}$ for the *I* images and 0.″1006 ± 0.″0003 pixel$^{-1}$ for the narrow-band images. The read-out noise was 20 e$^{-}$.

The observing log is shown in Table 1. The telescope was slightly offset between exposures to minimize flat field errors. The FWHM of stellar images ranged from 0.″4 in the beginning of each of the two exposure sequences (one in each filter) to 0.″6 for the last image, degraded more by defocus than seeing. These images are among the best resolved long exposures ever obtained from the ground using a non-image-stabilizing system. On short exposures some images had 0.″33 FWHM showing the high optical quality and superb site of the telescope.

The images were bias subtracted and flat fielded with twilight flats. We found the response of the CCD camera to be linear to 1 % as a consequence of a change of amplifiers prior to our observations (Nilsson 1993). Thus no non-linearity correction was applied to the images (as has been customary with Stockholm CCD images cf. Kjeldsen 1992).

To avoid that any of our results could be affected by artifacts arising from defocus we only included frames showing absolutely no sign of defocus in the two combined images. These consisted of the first four 5903/88 frames (#1–4) and the first three *I* frames (#9–11) and had FWHM of 0.″40 and 0.″47, respectively (measured by PSF fitting as described below; the values are 0.″44 and 0.″49 respectively if measured by fitting a sum of two Gaussians to a nearby star, cf. Table 1). The combined images are shown in Fig. 1. Coadded images (#1–8 and #9–16) constructed from all the frames, including defocused images, thus only served as confirmation of the results reported hereafter. These had FWHM of 0.″46 (5903/88) and 0.″49 (*I*).

Photometry was performed using DAOPHOT II and ALLSTAR (Stetson 1987). The point spread function (PSF) was determined from a star 20.″4 north of the quasar (and comparable in magnitude), using a fitting radius of 3 pixels, a Penny analytical model (sum of elliptical Gaussian and Lorentzian with different position angles), plus a look-up table for the residuals. In the PSF photometry a fitting radius of 2.5 was used.

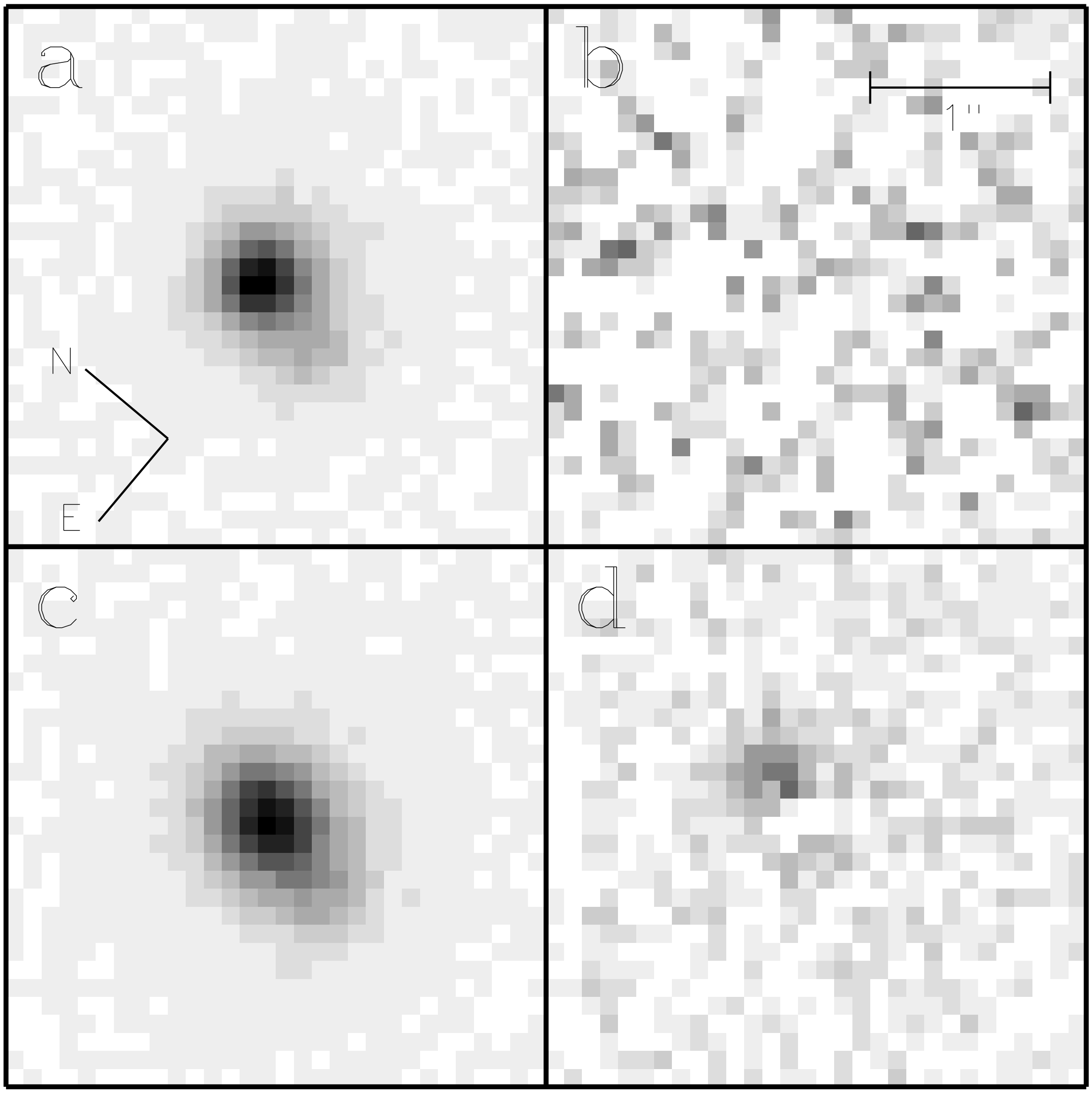

**Fig. 1.** Ly$\alpha$/N v and $I$ images of 1208+1011 at $\sim 0\farcs45$ FWHM resolution. **a** Excerpt of the 40 min. 5903/88 image giving a $3'' \times 3''$ view of the quasar. The A component is the bright northern mirage. **b** Result of PSF photometry of **a** using the nearby star as a model for the PSF. There are no significant residuals. **c** Same as **a** for the 15 min. $I$ image. **d** Result of PSF photometry of **c** using the nearby star as a model for the PSF. Notice the significant residuals. The images in **a** and **c** are scaled to equal maximum intensity, and the residuals in **b** and **d** have been scaled up by a factor of 5. A hard-copy of this figure (which is of higher quality) is available upon request (`jens@mail.ast.cam.ac.uk`)

### *2.1. Narrow-band image*

The image of the gravitational lens candidate in Ly$\alpha$/N v light is shown in Fig. 1a. The two components are clearly seen at $0\farcs4$ FWHM resolution. After the simultaneous fit of brightnesses and positions of the two components and subsequent double PSF subtraction no residuals were left above the noise, cf. Fig. 1b. The residuals are rather noisy, primarily because the PSF star used had very little intensity. We have therefore made experiments with reductions of simulated data (with realistic read-out noise and photon shot noise, but assuming perfect flat fielding) constructed to mimic the observed data (cf. Hjorth & Jensen 1993). From these simulations we can estimate the uncertainty in the derived intensity ratio. We find that the amplification of A relative to B is A:B=3.85±0.2.

The initial PSF photometry of the quasar (Fig. 1c) in the $I$ band gave almost the correct distance between A and B ($0\farcs46$) but the intensity ratio was found to be as low as A:B=3.0 and there were significant residuals left over after the double PSF subtraction (Fig. 1d). Furthermore, a MEM (Maximum Entropy Method) deconvolution (Hjorth & Jensen 1993) of the $I$ image revealed an elongated structure roughly on the line joining A and B and extending further out on both sides, with a length of $1''$. Finally, the entire quasar image is bright in $I$ ($m_I \approx 16\fm8$, cf. Table 2) making its colour quite red ($V - I \approx 1.1$) compared to other quasars at similar redshifts. This red colour is consistent with the spectrum of Sargent et al. (1986) from which we estimate the spectral index of the quasar spectrum above 6000 Å to be rather large ($f_\nu \sim \nu^\alpha$; $\alpha \approx -1.35 \pm 0.2$) compared to most high-redshift quasars for which $\alpha \approx -0.92 \pm 0.26$ (Schneider, Schmidt & Gunn 1991).

Combined these observations are indicative of a contribution from an underlying red object and could be the signatures of an intervening galaxy in the line of sight to the quasar, responsible for the gravitational lensing. However, reanalysis of the *HST* images revealed no such object (cf. § 3). Furthermore, although our images are found to be consistent with two point sources on top an exponential disk galaxy located $0\farcs1$ from A elongated in the direction of the line joining A and B, such a putative galaxy would have to be unusually bright ($I \sim 18$) for its suspected redshift. Thus, even though we cannot positively exclude the existence of a galaxy we find it more likely that the residuals are caused by PSF variation across the frame.

Gravitational lens studies have previously been hampered by artifacts in the photometry either as a result of detector non-linearities or due to PSF variation over the field, e.g., as a result of complicated optics (Magain et al. 1992b; Yee & Ellingson 1994). As we had verified that the CCD response was linear the most obvious systematic error would be a significant variation of the PSF over the field, in this case over a distance of $20''$. Photometry of the stars in the standard field in M92 (Christian et al. 1985)

**Table 2.** Photometry of the 1208+1011 system

| Object | $m_V$[a] | $m_R$[a] | $m_I$[b] |
|---|---|---|---|
| 1208+1011 | 17.91 | 17.38 | 16.75 |
| PSF star | 18.51 | 18.15 | 17.75 |

[a] From photometry of *NOT* images taken on June 22 1992 (Hjorth & Jensen 1993). The zero point of the absolute calibration is accurate to $\pm 0\fm07$

[b] From photometry of the *NOT* images taken on April 11 1993 reported in this paper. The zero point of the absolute calibration is only accurate to $\pm 0\fm2$

variation.

However, there cannot be any strong variation of the PSF from component A to component B of the quasar ($0\farcs47$). We therefore attempted an iterative determination of the PSF from the quasar itself to test the hypothesis that it consists of only two point sources. Thus, the B (A) component was subtracted with the position and brightness given by the first multiple fit, a new model for the PSF was determined from the remaining A (B) component, and the procedure was repeated until convergence (typically 2–5 times). A number of PSF models was constructed in this way. We found that in order to eliminate the residuals (basically, by absorbing them in the new PSF), the PSF had to have a FWHM of 7–10 % larger than that of the star. In the process of determining a useful PSF and for optimizing the fitting parameters the known separation between A and B was used as an indicator of a good fit. For example, solutions giving a separation of $0\farcs42$ was rejected. PSF photometry of the star gave some very conspicuous negative residuals, but the fit to the quasar was satisfactory yielding the correct separation of $0\farcs47$ and, by construction, almost no residuals. The intensity ratio was found to be A:B=3.35±0.2. Here the uncertainty reflects the systematic errors in the construction of the PSF. No acceptable PSFs, giving reasonable values for the image separation and PA of A and B, yielded intensity ratios outside the quoted interval. In particular, we were unable to construct a PSF yielding a value of A:B close to 4.

## 3. Discussion

We have performed a reanalysis of the publicly available archive *HST* images partly to look for the lensing galaxy and partly to check the intensity ratios in various pass-bands obtained at a different epoch. The images that we have reanalyzed are the ones previously studied by Bahcall et al. (1992a). They consist of eight Planetary Camera (PC) exposures, one F555W ($V$) 260 s exposure from December 23 1991 and one 600 s F555W exposure, two F439W ($B$) 900 s exposures, two F702W ($R$) 300 s exposures, and two F785LP ($I$) 400 s exposures from January 20 1992. The images were cosmic-ray cleaned by inter-comparison of each two images taken in the same filter and coadded to give four images, one in each pass-band.

The model for the PSF in each image was constructed in several steps. A F555W PSF was constructed using PC images of the quasars UM682, 0831+1248, and 1136+1214 taken in November 1991. These quasar images were situated within $1''$ from the position of 1208+1011 on the different CCD frames and were therefore expected to be good approximations to the true PSF. The constructed PSF was used as an initial guess for the PSF in the F555W image of 1208+1011 and was also used as a starting model for the F439W and F702W images. Similarly, a star in the globular cluster NGC 5904 (M5) observed in December 1991 was used as an initial guess for the F785LP PSF. A complication was that this star was saturated so that only the wings could be used.

**Table 3.** Intensity ratios and separations between A and B from *NOT* and *HST* data

| Telescope | Filter | Date | Ratio | Separation |
|---|---|---|---|---|
| | F439W | | $3.6 \pm 0.2$ | $0.''477 \pm 0.''004$ |
| | F555W | | $3.9 \pm 0.1$ | $0.''473 \pm 0.''002$ |
| | F702W | | $4.0 \pm 0.1$ | $0.''476 \pm 0.''002$ |
| *HST* | F785LP | January 20 1992 | $3.8 \pm 0.1$ | $0.''471 \pm 0.''003$ |
| | *V* | | $4.15 \pm 0.15$ | $0.''474 \pm 0.''008$ |
| *NOT*[a] | *R* | June 22 1992 | $3.95 \pm 0.15$ | $0.''470 \pm 0.''008$ |
| | 5903/88 | | $3.85 \pm 0.2$ | $0.''47 \pm 0.''01$ |
| *NOT* | *I* | April 11 1993 | $3.35 \pm 0.2$ | $0.''47 \pm 0.''01$ |

[a] From Hjorth & Jensen (1993)

Given a starting model for the PSF we then performed the first multiple fit of two point sources. No signs of a third component or a lensing galaxy showed up. Although this is not proof that the lensing galaxy is absent in the images it does support the interpretation of the *NOT* images discussed in the previous section. We therefore repeated the iterative determination of the PSF model from the images themselves *under the hypothesis that the images consist of two point sources* as detailed in § 2.2. The results are consistent with those found by Bahcall et al. (1992a) and are summarized in Table 3. The separation between A and B is found to be $0.''474 \pm 0.''003$ adopting a PC (chip 6) pixel scale of $0.''04374\,\mathrm{pixel}^{-1}$ (Gould & Yanny 1994), and their relative amplification is found to be almost perfectly achromatic. Analysis of the separate images (no coaddition) yielded similar results and enabled us to estimate the uncertainty in the parameters given in Table 3. The fact that our independent method of iteratively determining the PSF from the images themselves yields results similar to those of Bahcall et al. (1992a) (who used a different approach for determining the PSF) gives us confidence in the method and the results of § 2.

We thus find that the continuum intensity ratio in *I* (3.35) on April 11 1993 was smaller than what was observed on January 20 1992 with *HST* in F439W, F555W, F702W, and F785LP and on June 22 1992 in *V* and *R* (A:B=4±0.1 cf. Hjorth & Jensen 1993). Within the gravitational lens hypothesis there are essentially two possible physical processes that can give rise to such a time variation.

One of the possibilities is that the quasar is intrinsically variable. As the paths travelled by the two light rays of A and B are slightly different this gives rise to a time delay of one of the components. However, this explanation is not very likely because of the expected short time-delay in 1208+1011. A crude theoretical estimate of the time-delay is of the order of a few weeks (following Surdej et al. 1988).

The other possibility is that one of the components is being amplified through microlensing by individual stars in the intervening galaxy responsible for the splitting of the two images (macrolensing). The typical time scale for a microlensing event in 1208+1011 is of the order of a decade (Schneider et al. 1992) and thus appears to be a more likely explanation of the time variation.

There is additional evidence in support of this microlensing interpretation: Because of the very similar colours of A and B we can conclude that on April 11 1993 the continuum intensity ratio (3.35) was smaller than the emission-line intensity ratio (3.85). Microlensing in general amplifies the continuum emission, which arises from the more compact central regions of the quasar, more strongly than the light from the more extended emission-line regions. The larger intensity ratio that we see in the emission-line images compared to the continuum images may therefore indicate that component B is microlensed. (If A were the microlensed image we would expect to see the opposite effect, namely a larger intensity ratio in the continuum filter.) The observation of a larger emission-line intensity ratio also supports the indication seen in *HST* spectra (Bahcall et al. 1992b) that the blue wings of the emission-lines may be more strongly amplified than the continuum. A similar effect, attributed to microlensing, has been observed in e.g. the gravitational lens candidate HE 1104−1805 (Wisotzki et al. 1993). This interpretation is consistent with the simplest gravitational lens models (Refsdal & Surdej 1994) in which the smaller (weaker) component (B) is located close to the centre of the deflecting body where the cross section for microlensing is larger.

Thus, the time variation of the continuum intensity ratio and the difference in the continuum and emission-line intensity ratios in conjunction indicate that component B of 1208+1011 is being microlensed.

In this paper we have presented very high-resolution ground-based CCD images (FWHM $\approx 0''.4 - 0''.5$) of the gravitational lens candidate 1208+1011 obtained at the Nordic Optical Telescope. We have found that on April 11 1993 the emission-line (Ly$\alpha$/N V) intensity ratio between the two imaged components was 3.85±0.2 compared to an $I$ continuum intensity ratio of 3.35±0.2.

On the other hand our reanalysis of the available *HST* images from January 20 1992 confirms previous findings (Bahcall et al. 1992a) that the two components of 1208+1011, A and B, are separated by $0''.474 \pm 0''.003$, and have almost identical broad-band colours with an intensity ratio of about 4, also in the F785LP filter. This means that the small $I$ intensity ratio that we find in the *NOT* image probably cannot be explained as the reddening of the B spectrum by a foreground object, e.g., the deflecting galaxy. Neither have we found any convincing direct evidence for a lensing galaxy or a third lensed component, but it should be kept in mind that the above broad-band results were obtained assuming that the images should be accounted for by only two point sources.

If 1208+1011 is a gravitationally lensed quasar, a likely interpretation of the small $I$ intensity ratio (compared to previous broad-band data and the present narrow-band data) found in our *NOT* images is (i) that A:B has changed within the year 1992–1993, either due to a microlensing event or due to intrinsic variation of the quasar, leading to a time delay and (ii) that the Ly$\alpha$/N V emission-line is more strongly amplified relative to the continuum in the A component than in the B component, indicating that component B is microlensed. Another possibility, which cannot be excluded by the present observations, is that 1208+1011 is a binary quasar, in which case A and B may vary independently. The detection of a deflecting body between the two images or the measurement of a time delay, through monitoring of the light curves of the two components, would settle this question.

Finally, we would like to point out that the very high spatial resolution attainable without the use of adaptive optics (as evidenced by the observations reported in this paper) is an ideal starting point for the new high-resolution adaptive camera (HIRAC) under construction for the *NOT*. This camera is expected to reach angular resolutions of $0''.3$ FWHM on a regular basis and peak performances approaching the diffraction limit. The availability of such an instrument will be very beneficial for ground-based studies of gravitational lens systems e.g. for studies of physical properties of quasars, the distribution of dark matter in clusters of galaxies, gravitational lens statistics, and the determination of the Hubble constant. Future progress on 1208+1011 can also be made using high angular resolution ground-based imaging in the infrared. In addition, *HST* spectroscopy with higher S/N is essential to test the proposed microlensing hypothesis.

Arne Ardeberg for allocation of observing time. We are particularly indebted to Anlaug Amanda Kaas for her help during the preparatory phase of the observations and to Jean Surdej, Jean-François Claeskens, and Paul Schechter for very positive interaction and independent analysis of the data presented. We are also grateful to Michael Andersen, Marie-Christine Angonin-Willaime, Hans Kjeldsen, Geraint Lewis, Rolf Stabell, Anton Norup Sørensen, and Bjarne Thomsen for help, comments, and discussions. JH was supported by the Carlsberg Foundation, KN by the Academy of Finland, and LF by the Swedish Science Research Council (NFR).